\documentclass[useAMS,usenatbib]{mn2e}

\newcommand{\kms}{\mbox{km s$^{-1}$}}

\usepackage[hang, raggedright, tight]{subfigure}
\usepackage{graphicx}
\usepackage{amssymb}

\title[A Survey for Ground-State OH Masers Toward a Sample of Herbig-Haro Objects]{A Survey for Ground-State OH Masers Toward a Sample of Herbig-Haro Objects}
\author[A. de Witt, M. Bietenholz, R. Booth, M. Gaylard]{A. de Witt,$^{1}$\thanks{E-mail:
alet@hartrao.ac.za}  M. Bietenholz,$^{1,2}$\   R. Booth,$^{3}$\  and  M. Gaylard$^{1}$ \\
$^{1}$Hartebeesthoek Radio Astronomy Observatory, PO Box 443, Krugersdorp 1740, South Africa\\
$^{2}$Department of Physics and Astronomy, York University, 4700 Keele Street, Toronto, ON M3J 1P3, Canada\\ 
$^{3}$SKA South Africa, The Park, Park Road, Pinelands, 7005, South Africa}

\begin{document}

\date{}

\pagerange{\pageref{firstpage}--\pageref{lastpage}} 

\maketitle

\label{firstpage}

\begin{abstract}
Herbig-Haro objects are regions of shocked gas and dust which are
produced when collimated outflows from a protostar interact with the
surrounding dense gas.  They have many similarities to
supernova remnants which are interacting with molecular clouds.
1720-MHz OH masers have been identified towards a number of
interacting supernova remnants.  Observations and models indicate that
these masers are shock excited and are produced behind C-type shocks.
If conditions behind the shock fronts of Herbig-Haro objects are similarly able to support 1720-MHz OH masers they could be a useful diagnostic tool for star formation.  We therefore searched for 1720-MHz OH maser emission towards a sample of  97 Herbig-Haro objects using the Green Bank radio
telescope. We detected 1720-MHz OH lines in emission in 17 of them, but neither their spectral signature nor follow-up observations with the Very Large Array  showed any conclusive evidence of maser emission.  We conclude that the emission detected from our single-dish
observations must be extended and most likely originates from
thermal or quasi-thermal excitation processes. We also investigated the properties of Herbig-Haro shocks more closely and conclude that despite the overall similarities to supernova remnants,  the conditions required for maser emission, in particular,  a sufficient velocity-coherent column density,
 are not likely to occur in Herbig-Haro objects. 
 
 \end{abstract}


\section{Introduction}
Star formation is accompanied by the ejection of mass from protostellar sources,  which is often seen in the form of collimated outflows. Where the overlying extinction is low enough to permit optical observations, structures called Herbig-Haro (HH) objects have been observed towards such outflows, mainly (but not exclusively)  from  protostars with $M_{\star} \leq M_{\odot}$.  HH objects are  regions of shocked gas and dust that show the locations of impact between protostellar outflows and their surrounding environments. These powerful outflows are usually in the form of jets and the HH objects are  sometimes observed  as isolated entities and sometimes as a collimated chain of knots along the jets.  The emission-line properties of  HH objects imply that the gas is excited by shock waves \citep{Schwartz1975, Schwartz1978}, and the characteristic low-excitation emission lines in the optical spectral range set them apart from the well known H\footnotesize{\MakeUppercase{\romannumeral 1}\MakeUppercase{\romannumeral 1}} \normalsize regions associated with high-mass protosteller objects. Observations of HH objects show that many of these outflows are interacting with dense molecular material and have  been found to be  accompanied by  strong, near-infrared (NIR) lines from the H$_{2}$ molecule, at 2.12 $\mu$m.  It is widely recognised that the NIR emission lines of the H$_{2}$ molecule at 2.12 $\mu$m are collisionally excited in slow, non-dissociative shocks and are one of the primary signatures of C-type shocks in dense molecular material (\citeauthor{Draine1982} \citeyear{Draine1982}; \citeauthor{Draine1983} \citeyear{Draine1983}; \citeauthor{DeNoyer1983} \citeyear{DeNoyer1983}). HH objects are sometimes seen as bow-shaped  structures that show signatures of J-type shocks (discontinuous, dissociative shocks) at the apex and C-type shocks (continuous, non-dissociative shocks) in the wings of the bow. For example,  \citet{Smith2003} used modelling of HH7 to successfully explain their NIR observations of the 2.12  $\mu$m,  H$_{2}$ line emission in terms of a C-type shock on the wings of a bow-shaped structure. 

In recent years weak masers from one of the ground-state lines (at 1720 MHz) of the OH molecule have been found to be associated with a sample of supernova remnants \citep[SNR; e.g.\ ][]{Green2002}. These masers are believed to be collisionally excited behind the shock front where the ejected material from a supernova explosion rams into the interstellar medium (ISM). 
The inferred interactions have been confirmed by follow-up searches for millimetre or IR emission from hot molecular gas or from molecules produced by the rich chemistry occurring within the shock front (e.g.\ \citeauthor{Frail1998} \citeyear{Frail1998}; \citeauthor{Reynoso2000} \citeyear{Reynoso2000}; \citeauthor{Zhou2009} \citeyear{Zhou2009}).   To date, 1720-MHz OH masers have been found in about 10\%  of the more than 250  SNRs searched (e.g.\ \citeauthor{Yusef-Zadeh1995} \citeyear{Yusef-Zadeh1995}; \citeauthor{Green1997} \citeyear{Green1997}; \citeauthor{Brogan2004} \citeyear{Brogan2004}; \citeauthor{Hewitt2009} \citeyear{Hewitt2009}). 
 \citet{Lockett1999},  modelled these regions  and showed that under a restricted set of physical conditions, 1720-MHz OH masers can be collisionally excited behind a C-type shock formed when a SNR interacts with a dense molecular cloud within the ISM.  NIR emission from H$_{2}$ at  2.12  $\mu$m is often seen to extend over large regions of a SNR, with compact maser spots located near the emission peaks. Based on the tight constraints under which the  masers in SNRs form, \citet{Lockett1999} suggested that the 1720-MHz OH masers could be an important signpost of C-type shocks and the interaction with dense molecular material. 

HH objects, like supernova remnants, also contain shocks.  The shocks in HH objects occur in a wide range of physical conditions, and it is therefore possible that some of the mechanisms that operate in SNRs would also operate in HH objects, and that 1720-MHz OH masers could also be excited in HH objects.  However, observations to detect 1720-OH masers have rarely been undertaken towards HH objects, and there have been no blind searches of the Galaxy for 1720-MHz OH masers.  It is therefore possible that OH masers associated with HH objects are common but have not so far been detected.

We therefore undertook a large survey of HH objects to search for OH maser emission using the Green Bank Telescope (GBT), followed by Very Large Array (VLA) observations of the more promising of the GBT detections. Generally, single-dish observations of the 1720-MHz OH masers associated with SNRs show  weak, narrow lines, with broad-line absorption features at 1665 and 1667 MHz.  Therefore, the detection of emission lines at 1720 MHz with no corresponding 1665- and 1667-MHz emission would be a strong indicator of maser emission. Generally, maser emission occurs in very compact spots and has high brightness temperatures. Interferometric observations are generally required to confirm these last two characteristics.

In this paper, we present the results of these observations, and discuss their implications.

\section{Observations and Results}
\subsection{Source Selection}

As mentioned, only 10\% of SNRs searched have 1720-MHz OH masers.  If a similarly small fraction of HH objects have OH masers, then a large number of sources need to be surveyed to find at least a few detections of maser emission.  Since shock-excited masers are not necessarily strong,  a survey with high sensitivity is required. An efficient way to search a large number of sources for a particular, weak astrophysical feature is to make the first observations with a single-dish telescope of moderate resolution and high sensitivity. For this reason our survey is composed of two parts, firstly identifying OH line emission towards HH objects by doing single-dish observations, followed by interferometric observations to confirm the presence of maser spots.  This procedure was used with great success in surveys for 1720-MHz OH masers towards SNRs (\citeauthor{Yusef-Zadeh1995} \citeyear{Yusef-Zadeh1995}; \citeauthor{Frail1996} \citeyear{Frail1996}; \citeauthor{Yusef-Zadeh1996} \citeyear{Yusef-Zadeh1996}; \citeauthor{Green1997} \citeyear{Green1997};  \citeauthor{Koralesky1998} \citeyear{Koralesky1998}; \citeauthor{Yusef-Zadeh1999} \citeyear{Yusef-Zadeh1999}).

We selected a sample of 97 objects from the $\sim$400 listed in Reipurth's 1999 catalogue\footnote{Reipurth B., 2000, VizieR Online Data Catalog, 5104} of HH objects. Because the OH masers are likely to be weak, we have chosen the nearest objects, which have the best chance of being detected, from Reipurth's catalogue as primary targets for our initial survey. 

\subsection{Single-Dish Observations}
Observations using the Greenbank $\sim$105~m telescope at 1720, 1665 and 1667 MHz were made on 2005 February 05 and 07\@. The quadrants of the spectral correlator were configured for the OH ground-state lines at 1720  and 1667 MHz in left
and right circular polarization (LCP and RCP; IEEE convention).  A bandwidth of 12.5 MHz provided a resolution at 1720 MHz of 1.6 kHz per channel, corresponding to a velocity coverage of more than 2000 \kms\ at a velocity resolution of 0.13 \kms.  The total integration time was 15 min per source, giving a typical noise level of $\sim30$ mJy. Observations were done in frequency switching mode using half the integration time of 15 min per source for the nominal frequency and the other half shifted by 1 MHz.  This corresponds to a velocity shift of 349 \kms\ at 1720 MHz.  Sometimes more than one object fell within the 7\arcmin\ beam of the telescope so that the number of objects observed (97) exceeds the number of pointings (45)\@. The coordinates of the pointing centres as well as the HH objects that fell within the field-of-view for each of the observations are presented in Table~\ref{table1}.  For each of the observations, the object of most interest has been chosen to be at or close to the pointing centre and those objects are listed in boldface. 
In this paper, we refer to specific GBT observations by the name of
the object at or closest to the pointing centre, even though other
objects may also be visible.

The GBT spectra were converted to SDFITS format at Greenbank and processed further using NRAO's GBTIDL\footnote{GBTIDL: Data Reduction for the GBT Using IDL, http://gbtidl.nrao.edu/} software. The data were folded and calibrated to Janskys. Linear baselines were fitted over selected spectral regions where lines occurred and subtracted from the data so that the zero-level for line profiles could be found. LCP and RCP data were analysed separately. Where lines are present in the spectra, Gaussian profiles were fitted by least squares, using the interactive FITGAUS routine, and the parameters for the line profiles tabulated. In the case of polarised data the LCP and RCP lines were fitted separately and when no significant polarisation was seen, the LCP and RCP data were combined. The fitted Gaussians were subtracted from the data, and in no case were residuals larger than expected from the noise.

\begin{table*}
 \centering
 \begin{minipage}{140mm}
  \caption{HH objects selected for observations with the Greenbank Telescope.}
  \label{table1}
  \begin{tabular}{@{}llllr@{}}
  \hline
   RA (J2000)     		&       DEC (J2000)    		& Objects contained in pointing\footnote{{A list of all HH objects that fell within the GBT field-of-view for each observation. For each of the observations the object of most interest has been chosen to be at or close to the pointing centre and is listed in boldface.}}   & Region & Distance            \\
  (dd mm ss.s)          &       (dd mm ss.s)	&        & 		 &  (pc)          \\

 \hline
00 37 13.99    &$+$64 04 17.0   &\textbf{HH288}                                  				&Cassiopeia     	&2000 \\
03 01 32.70    &$+$60 29 12.0   &\textbf{HH163}                                 				&IC 1848A       		&2200 \\
03 27 18.66    &$+$30 17 20.1   &\textbf{HH279}                                  				&L1455         		&300 \\
03 28 49.69    &$+$31 01 13.4   &\textbf{HH14}                                   				&NGC 1333       	&220 \\
03 28 57.56    &$+$31 20 08.0   &HH6/\textbf{12}/334                             			&NGC 1333       	&220 \\
03 28 58.94    &$+$31 08 00.3   &HH13/\textbf{15}/16/341/342/343/350                 	&NGC 1333       	&220 \\
03 29 08.10    &$+$31 15 26.4   &HH5/6/\textbf{7}/8/9/10/11/344/345/346/34       	&NGC 1333       	&220 \\
03 29 20.36    &$+$31 12 50.2   &\textbf{HH5}/345/346/347/348/349                           &NGC 1333       	&220 \\
03 29 26.38    &$+$31 07 38.4   &\textbf{HH18}                                   				&NGC 1333       	&220 \\
03 43 56.19    &$+$32 00 51.9   &\textbf{HH211}                                  				&IC 348         		&300 \\
04 14 17.09    &$+$28 10 58.9   &\textbf{HH220}                                  				&L 1495     		&140 \\
04 18 51.39    &$+$28 20 25.8   &\textbf{HH156}                                  				&L 1495    		&140 \\   
04 22 00.70    &$+$26 57 42.2   &\textbf{HH157}/276                              			&L 1495  			&140 \\
04 27 01.77    &$+$26 05 43.6   &HH158/\textbf{159}                              			&Taurus   			&140 \\
04 28 17.93    &$+$26 17 46.7   &\textbf{HH31}                                   				&B218           		&140 \\
04 38 28.15    &$+$26 10 54.7   &\textbf{HH230}/233                              			&Taurus     		&140 \\
04 42 37.29    &$+$25 15 45.7   &\textbf{HH231}                                  				&Taurus     		&140 \\
05 07 49.51    &$+$30 24 07.0   &\textbf{HH229}                                  				&RW Aur jet     		&140 \\
05 34 11.38    &$-$06 34 00.1   &\textbf{HH84}                                   				&L1641          		&470 \\
05 35 17.81    &$-$06 17 42.3   &\textbf{HH33}/40/85/322/323                           	 	&L1641          		&470 \\
05 35 29.91    &$-$06 27 00.5   &\textbf{HH34}/134                               			&L1641          		&470 \\
05 36 00.15     &$-$06 35 59.4   &\textbf{HH86}/87/88/173/327                            		&L1641         		&470 \\
05 36 05.23    &$-$05 03 44.2   &\textbf{HH128}                                  				&L1641          		&470 \\ 
05 36 17.23    &$-$06 42 27.5   &HH1/\textbf{3}/35/144/145/147/148                          	&L1641          		&470 \\
05 36 21.23    &$-$06 46 07.4   &HH1/2/\textbf{144}/145/146/147/148                         &L1641          		&470 \\
06 39 05.82    &$+$08 51 26.7   &\textbf{HH39}                                   				&R Mon          		&800 \\
06 41 02.46    &$+$10 15 04.9   &\textbf{HH124}                                  				&NGC 2264       	&800 \\
18 29 18.89    &$+$01 14 17.5   &\textbf{HH106}                                  				&Serpens        		&310 \\
18 29 48.12    &$+$01 25 57.4   &\textbf{HH107}                                  				&Serpens        		&310 \\
18 35 36.14    &$-$00 35 09.3   &\textbf{HH108}/109                              			&Serpens        		&310 \\
18 54 05.77    &$+$00 32 40.4   &\textbf{HH172}                                  				&Aquila         		&500 \\      
19 17 57.36    &$+$19 11 56.4   &\textbf{HH223}                                  				&L723           		&300 \\
19 36 51.62    &$+$07 34 10.2   &\textbf{HH119}                                  				&NGC2264/B335   	&250 \\
20 45 53.74    &$+$67 57 43.4   &\textbf{HH215}/415                              			&L1158          		&500 \\
20 55 10.34    &$+$77 31 26.9   &HH199/\textbf{200}                              			&L1228          		&300 \\
20 59 11.96    &$+$78 22 53.4    &\textbf{HH198}                                  			&Cepheus       		&300 \\
21 42 26.81    &$+$66 04 27.8   &HH103/\textbf{232}/237/238/239/242                       &NGC 7129       	&1000 \\
21 42 40.29    &$+$66 05 41.6   &HH103/167/232/236/237/\textbf{238}/239/242       &NGC 7129       	&1000 \\
21 43 06.63    &$+$66 06 54.7   &HH105/\textbf{167}/234/236/237/238/239/242     	&NGC 7129       	&1000 \\
21 43 29.70    &$+$66 08 38.1   &\textbf{HH234}/105/167/235/236                         	&NGC 7129       	&1000 \\
22 19 38.65    &$+$63 32 38.3   &HH251/\textbf{252}/253/254                         		&S 140          		&900 \\
22 35 24.42    &$+$75 17 06.3   &\textbf{HH149}                                  				&L 1251         		&300 \\
22 38 38.83    &$+$75 10 37.3   &\textbf{HH189}/364                              			&L 1251         		&300 \\
22 56 03.54    &$+$62 02 01.3   &\textbf{HH168}/169                              			&Cep A          		&700 \\
22 56 58.53    &$+$62 01 42.4   &HH169/\textbf{174}                              			&Cep A          		&700 \\
\hline
\end{tabular}
\end{minipage}
\end{table*}

\subsection{Single-Dish Results}
The spectra from our GBT observations show detections for all but 4 of the observed sources,  HH163, HH229, HH128, and HH39, for which no  emission or absorption lines were detected at any of the observed frequencies.  Three of the sources, HH168 and HH174 towards Cepheus A, a well studied star-forming region, and  HH172 towards Aquila show clear evidence of previously detected OH main-line masers associated with H\footnotesize{\MakeUppercase{\romannumeral 1}\MakeUppercase{\romannumeral 1}} \normalsize regions \citep[e.g.\ ][]{Cohen1990, Szymczak2004}, and these sources will not be discussed further in this paper.  We note, though, that HH172 also showed absorption at all of the observed frequencies at a local-standard-of-rest (LSR) velocity distinct from the detected maser emission. 
The spectra for the remaining sources all show both the 1665- and
1667-MHz OH lines in emission, and for the 1720-MHz line, show
examples of pure noise, pure absorption, pure emission and, in a few
cases, both emission and absorption.

The 1720-MHz line is seen in emission in 17 of our spectra and absorption lines were detected in 4 spectra, while no 1720-MHz OH lines were detected in 14 of the spectra. The  remaining 5 spectra show a combination of both emission and absorption lines, with a continuous change from emission to absorption, with the emission at a higher frequency than the absorption in two of the spectra and vice versa for the other three spectra. 

The observed emission lines from all transitions were weak. With the exception of a few sources the spectra exhibit symmetric, single feature profiles with no significant  linear polarisation and line widths broader than that expected for  a single maser spot.   Line widths for the emission lines ranged between 0.25--3.2 \kms\ and peak flux densities range between  0.02--0.81 Jy.  Apart from HH288 all of the spectra have peak LSR velocities between $\sim -10$ and +12 \kms.

Although we detected 1720-MHz OH emission for a number of sources in our GBT observations, none of them show unambiguous 1720-MHz OH
maser emission, i.e.\ 1720 MHz in emission with 1665 and 1667 MHz in
absorption or absent.  The 1720-MHz emission and absorption lines are
mostly weak compared to the main lines with line widths and peak line
velocities similar to those of the 1665- and 1667-MHz emission lines.

Masers from SNRs are generally weak and in single-dish spectra the spatial blending of compact maser components often makes it difficult to differentiate between maser emission and the widespread Galactic thermal OH emission at 18 cm, which was first detected by  \citet{Turner1973} and which is associated with giant molecular clouds. Observations have established that away from intense IR emission sources,  the widespread thermal OH emission detected by \citet{Turner1973} has weak, single-line profiles with broad, $\sim$5 \kms,  line widths and  main-line (1667 MHz/1665 MHz) intensity ratios characteristic of local thermodynamic equilibrium (LTE), between 1 and 1.8,  ranging from optically thin to optically thick conditions.   Anomalies in the main-line intensity ratios and relative line shapes occur for only a few cases, and anomalies where the 1665-MHz line is too strong relative to the 1667-MHz line appear to be uncommon,  while the 1720-MHz line almost always shows anomalous emission. Most of the spectra from our GBT observations display such thermal behaviour  with near equilibrium intensity ratios of the main-lines, although the line profiles are narrower,  $<$ 3 \kms, than those detected by \citet{Turner1973}.  However,  OH spectra with  thermal line ratios  and narrow line widths,  $<$ 2 \kms,  are  commonly detected towards dark clouds with modest opacity (e.g.\ \citeauthor{Slysh1994} \citeyear{Slysh1994}; \citeauthor{Li2003} \citeyear{Li2003}), and such clouds are commonly associated with HH objects.   

\begin{figure*} 
\centering
\begin{minipage}{140mm}
\includegraphics[width=140mm]{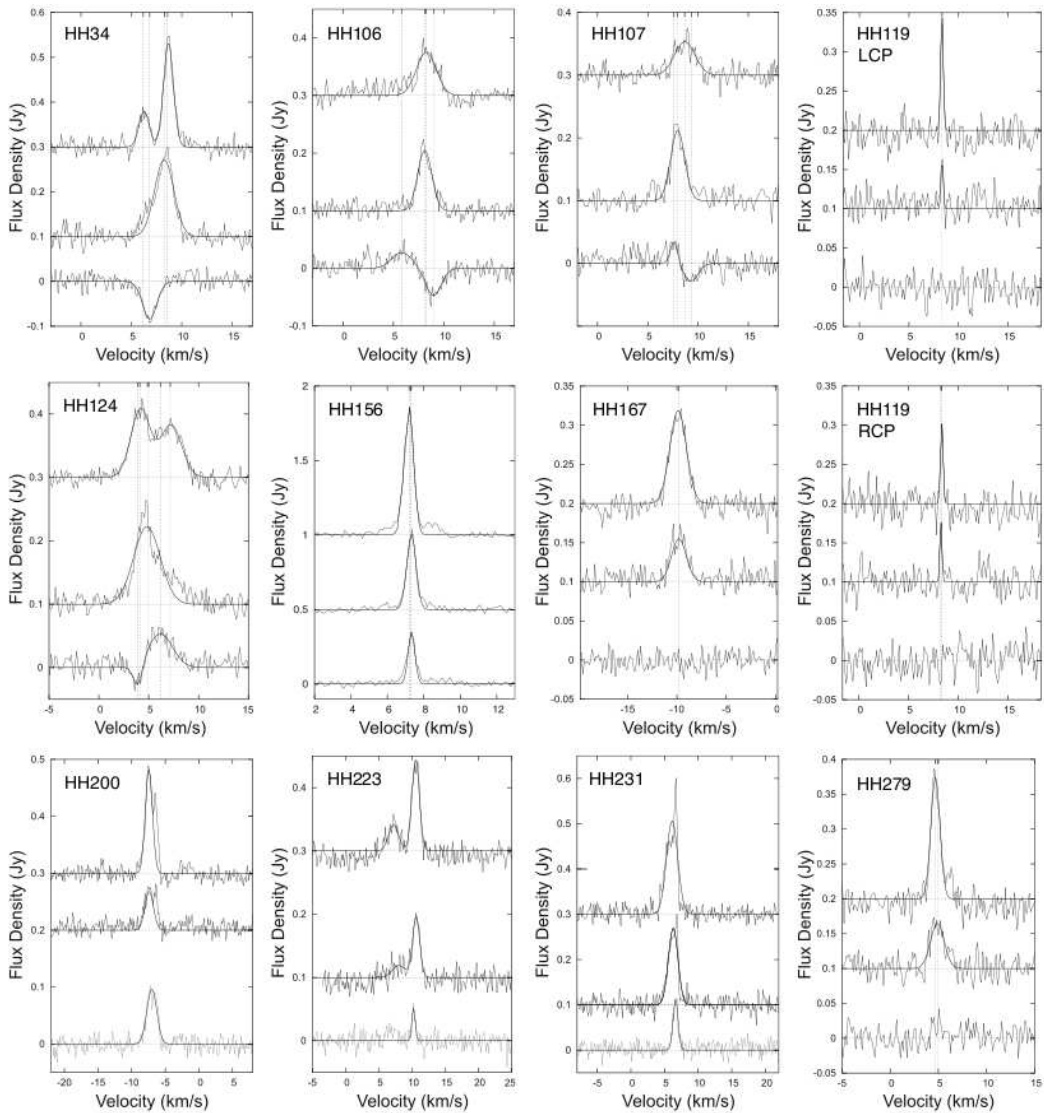}
\caption{OH spectra showing possible maser emission obtained from the GBT observations. Top: 1667 MHz, Middle: 1665 MHz, Bottom: 1720 MHz. The solid curves are the fits of the 1-D Gaussians to the spectra and the dotted lines indicate the peaks of the Gaussians fitted to the  emission and absorption lines.}
\label{figure1}
\end{minipage}
\end{figure*}

While the majority of our sources show spectra indicative of thermal emission,  we did identify 11 objects with line profiles that appear to be dominated by non-thermal excitation mechanisms.  One source,  HH231, shows signatures of non-thermal emission in the 1720-MHz line and a further 10 sources show anomalous emission in the main lines, that could be maser emission. These sources were identified as having anomalous emission because they  exhibit one  or more of the following properties; non-characteristic LTE line ratios, double-peaked emission lines and strong, narrow lines compared to the other sources in the sample. HH119 in addition shows what appears to be a slight difference between the LCP and RCP spectra.   The observed spectra are presented in Figure~\ref{figure1} and the Gaussian fits to the spectra are listed in Table~\ref{table3}.  

\begin{table*}
 \centering
 \begin{minipage}{140mm}
  \caption{Parameters of the 1665, 1667 and 1720 MHz line profiles of
    the spectra identified as maser candiates, found by fitting
    Gaussians to the GBT spectra.  We list the spectra by the name of the
    HH object at or closest to the GBT pointing centre, although additional HH objects may be within the 
    telescope beam, see Table~\ref{table1}.  We list
    the velocity, FWHM and intensity of the peak.}
  \label{table3}

  \begin{tabular}{@{}lrrrrrrrrr@{}}
  
\hline
	     & \multicolumn{3}{l}{\textbf{1665 MHz Parameters}} & \multicolumn{3}{l}{\textbf{1667 MHz Parameters}} & \multicolumn{3}{l}{\textbf{1720 MHz Parameters}} \\
 & Velocity & Width & Peak   & Velocity & Width & Peak  & Velocity & Width & Peak \\
Pointing\footnote{{The HH object name listed refers to a specific pointing centre as referred to in Table~\ref{table1}.}}             & (\kms) &(\kms) &(Jy) & (\kms) & (\kms) & (Jy) & (\kms) & (\kms) & (Jy) \\

\hline
\textbf{HH34}          &	8.30	 	&2.25        &0.17	&8.59	  	&1.16	&0.23       &$-$           &$-$          &$-$\\
                                  &$-$             &$-$          &$-$          & 6.18                 & 1.18      & 0.08       &6.82	&1.64	&-0.08\\
\textbf{HH106}        &	8.11		&1.65	&0.10        &8.26		&2.31	&0.08	&8.98	&1.78	&-0.05\\
			      &$-$		&$-$		&$-$		&$-$			&$-$		&$-$		&5.83	&2.36	&0.03\\
\textbf{HH107} 	      &	7.98		&1.64	&0.11	&8.67		&2.39	&0.05	&7.61	&0.85	&0.04\\
			      &$-$		&$-$		&$-$		&$-$			&$-$		&$-$ 	&9.32	&1.60	&-0.03\\
\textbf{HH119-lcp} &	8.35	 	&0.25	&0.07	&8.34		&0.32	&0.15	&$-$		&$-$		&$-$ \\
\textbf{HH119-rcp} & 8.26 	& 0.25	& 0.08	& 8.36		& 0.31 	& 0.10 	& $-$	&$-$		&$-$ \\
\textbf{HH124}        &	4.87		&3.14	&0.12 	&4.10		&2.29	&0.11	&3.91	&1.10	&-0.04\\
			      &$-$		&$-$		&$-$ 	&7.16		&2.50	&0.08	&6.18	&2.87	&0.05\\
\textbf{HH156}        &	7.27	         &0.60        &0.51        &7.19                  &0.62	&0.81	&7.31        &0.59	&0.32\\
\textbf{HH167}        &	-9.80	 	&1.75	&0.05	&-9.88		&1.98	&0.12        &$-$		&$-$		&$-$ \\

\textbf{HH200}        &$-$7.36	&1.26	&0.07	&$-$7.43		&1.19	&0.19	&$-$7.33	&1.09	&0.09\\
                            	      &$-$6.36	&0.55	&0.06	&$-$6.38		&0.53	&0.12	&$-$6.41	&0.76	&0.07\\

\textbf{HH223}        &	7.24		&2.08	&0.03	&7.78		&2.30	&0.02	&6.98	&1.85	&0.02 \\
			      &	10.58	&1.38	&0.13	&10.61		&1.11	&0.10	&10.24	&0.49	&0.05\\
			      
\textbf{HH231}        &	5.78		&1.81	&0.17	&6.07		&1.55	&0.15	&6.64	&0.89	&0.11\\
			      &	6.69		&0.44	&0.21	&6.81		&0.36	&0.12	&$-$		&$-$		&$-$\\

\textbf{HH279}        &	4.89		&1.68        &0.06	&4.69      		&1.11	&0.17	& $-$	&$-$		& $-$\\

\hline
\end{tabular}
  \end{minipage}
\end{table*}

\subsection{VLA Observations}
Our choice of objects to observe with the VLA was based on the GBT observations, but was
constrained by the total amount and LST range of the VLA time
obtained.  We were able to observe 5 of the 11 sources for which our
GBT observations suggested non-thermal excitation. We obtained high-resolution VLA observations of these five sources to confirm whether compact maser spots were indeed present.  The five candidates were  HH106, HH107, HH167, HH200 and HH279. The VLA observations were made in four sessions  on 2007 July 15 \& 16 and  August 5 \& 12. 
We spent one hour on-source for each of HH167, HH200 and HH279, while HH106 and HH107 were observed together for one hour on-source by pointing midway between them, as both fell within the VLA field-of-view.  Details of the VLA observations and  observational setup are presented in Table~\ref{table2}.

\begin{table*}
 \centering
 \begin{minipage}{140mm}
  \caption{HH objects selected for observations with the VLA.}
  \label{table2}

  \begin{tabular}{@{}llcllrrlr@{}}
  
  \hline
Observing &  HH & Time on  & OH lines & Peak  & Expected\footnote{The expected SNR, assuming the flux density measured from the GBT observations are unresolved. The expected SNR was estimated from nominal sensitivity of the VLA and scaled for the actual number
of visibilities and for a channel width of 6.104 kHz.}  & RMS\footnote{The measured background rms per channel as obtained from the central channel of the dirty image.}  & Beamsize\footnote{The FWHM of the fitted Gaussian restoring beam}, P.A.  & T$_{B}$\footnote{Peak observed brightness temperature in the image, calculated using the FWHM of the fitted Gaussian restoring beam and the maximum pixel  brightness from the dirty image}\\
Date & object & source (min) & MHz & mJy & SNR &  mJy/beam  &   arcsec $\times$ arcsec, deg  & K\\

 \hline
15/07/2007		&HH106/ 		& 35		&1665	&64.91		&13 & 10.64 &3.08 $\times$ 1.14, -49.53 & 8149\\
				& 107		&    		         &1667       &61.44	 	&10 &9.91 &2.19 $\times$ 1.16, -51.23 & 10661\\
				&		         &            	         &1720       &57.10		&7 & 9.36	&2.17 $\times$ 1.11, -46.36 & 9765\\	
				
16/07/2007           &HH279		& 38   		&1665    	&59.85 		&9 &8.53 &1.90 $\times$ 1.45, 42.93 & 9576\\
				& 		      	&   		         &1667   	&59.00		&24 & 9.09	&3.37 $\times$ 1.59, 17.56 & 4854\\
				
05/08/2007	&HH167	          &35		&1665    	&41.26   		&8 & 6.16   &1.47 $\times$ 1.20, -45.53 & 10311\\
				& 		          &          		&1667    	&37.90		&20 & 6.23 &1.52 $\times$ 1.17, -43.57 & 9394\\
					
12/08/2007		&HH200	        	&36		&1665       &55.72	 	&9 &9.57  &1.63 $\times$ 1.23, -37.20 & 12251\\
				& 		         &          		&1667    	&83.99   		&25 & 13.53 &1.66 $\times$ 1.21, -41.83 & 18433\\
				&		        	&                     	&1720    	&55.03		&16 & 9.04 &1.65 $\times$ 1.20, -39.45 & 11448\\	
\hline
\end{tabular}
\end{minipage}
\end{table*}

The observations were made using the A array configuration of the VLA.  The sources were observed at L band in either 2-IF or 4-IF mode with a bandwidth of  0.195 MHz per IF and 128 channels, using two centre frequencies per IF pair at both senses of circular polarisation.  We observed 1665 MHz and 1667 MHz together in 2-IF mode for those sources for which no 1720 MHz emission was seen in the single-dish data, and in 4-IF mode we observed 1665 MHz and 1720 MHz together and 1667 MHz and 1720 MHz together.  Our target sources have central velocities with respect to the LSR of between 5 and 10~\kms\ and line widths $\Delta v \leq 2.2$ km s$^{-1}$. We used the smallest available bandwidth of $\Delta \nu = 0.195$ MHz which at L band corresponds to a velocity range of 34~\kms\ which is sufficient for the target sources, and provides a resolution of 1.526 kHz per channel corresponding to a velocity resolution of 0.27~\kms.  

Observations were made using the  then new Expanded-VLA (EVLA) computer system controlling the array,  and nine operational EVLA antennas were included in the array by default. At that stage the VLA was undergoing a significant upgrade to become the EVLA,  and the EVLA antennas were offered on a shared-risk basis. Unfortunately, we had to discard all data from the EVLA antennas, as it proved unusable for our observations due to lack of phase stability. The shortest spacing between antennas that were actually used was 2.1 K$\lambda$ at 1720 MHz.

Calibration and imaging were done in the Astronomical Image Processing System (AIPS). For each of the datasets, the frequency bands at 1720, 1665 and 1667 MHz,  were calibrated separately.  The flux density of the primary calibrator, was  calculated for our frequencies of interest from the well-determined spectrum of \citet{Baars1977}.  The primary flux calibrators used were 3C48, 3C147 and 3C286\@.  Since the flux-density calibrators that we used are somewhat resolved, we used models of their source structure which are distributed with AIPS to determined antenna-based complex gain solutions from the visibility measurements.  Since the secondary calibrators are known to be unresolved at our wavelengths and resolutions, we used a point source model to solve for complex antenna gains from the visibility measurements of the secondary calibrator, boot-strapping the flux density scale from the solutions obtained for the flux-density calibrator.  We determined the  bandpass response also from the observations of the primary calibrators.

\subsection{VLA Results}
We produced multi-channel images or image-cubes for all of our target sources and for each of our frequency bands. No obvious emission was seen in any of these images, and so we did not do any deconvolution (CLEAN'ing), and our results  are based on the dirty images. For each of the  image cubes we used 113 channels, discarding 15 noisy channels from  the edges of the band,  and a strip of  50 pixels around the edges of the cube to avoid edge effects. 
We then obtained  the positive and negative extrema as well as the rms pixel brightness, $\sigma$,   for each of the image cubes using the AIPS task IMSTAT, to determine whether any real emission is present in the image cubes.  In no case were brightnesses in excess of $6.2 \sigma$ seen, apart from an unrelated continuum source in the HH279 field.  Also in no case, except for the continuum  source in the HH279 field, was the  magnitude of the positive extremum larger by more than $0.3\sigma$ than that of the negative one.  Since the negative extremum is likely due to noise, this implies that there is no significant emission even if the distribution of brightnesses deviates from a Gaussian one, provided it remains symmetrical about zero.  Furthermore the extrema, again with the exception of the continuum source in the HH279 field, occurred at different locations in adjacent channels, while for a real source one would expect the emission to be spread over a few channels, and thus for maxima to appear at the same location in adjacent channels.

The biggest, full-resolution,  image cubes that we made were 4096 $\times$ 4096 pixels, with a pixel-size of 0.4\arcsec,  thus having a width comparable to the VLA field-of-view.  Searching the whole VLA
field of view not only ensured that we covered an area greater than
the GBT position uncertainty, but also allowed us to include
a number of additional HH objects which were outside the
GBT beam.

We first average in frequency to obtain channels of width
6.104 kHz or 1.1 \kms\ in order to get a frequency resolution
comparable to the linewidths seen at GBT.   We detected no obvious
sources in either emission or absorption above the noise apart from the continuum source in the HH279 field.  We give
the brightness maxima and rms values in Table~\ref{table2}. We then further averaged in frequency to obtain channels of width 12.208 kHz or  2.2 \kms, comparable to the maximum line widths seen at GBT. Finally, we averaged all the channels together in order to look for any continuum sources in the field-of-view that could influence the results. In none of those cases did we detect any sources apart from the continuum source in the HH279 field.

In order to look for any features with narrow line-widths, we then also searched the cubes with our full frequency resolution
of 1.526 kHz (0.27 \kms), and also detected no significant emission apart from the continuum source. As a consistency check we also visually inspected the region and channel corresponding to the positive extrema in each of the image cubes,  as well as the channels where we expected to see emission from the LSR velocities obtained form our single-dish observations.  We did not find any significant emission in any of the fields  at any of the frequency bands observed.

To increase our sensitivity to more extended sources we also
made lower resolution images by down-weighting the longer baselines,
resulting in beam sizes of $\sim$3\arcsec.  In none of these lower-resolution image cubes did we see any significant emission except for the continuum source already mentioned.  Our listed upper limits
were obtained using natural weighting for the highest sensitivity,
however, we also made image-cubes using uniform weighting to minimise
the sidelobes, and no significant emission was seen on these either.

In the continuum image for the HH279 field, we did detect NVSS J032628+301618 at both 1665 and 1667 MHz, the only detectable NVSS source which fell within our fields of view.  We measured a flux density of $64 \pm 5$~mJy for it (after correction for the primary beam response), a value in reasonable agreement with the NVSS flux density of 72 mJy \citep{Condon1998} given that the source, like most extragalactic compact radio sources, is probably variable.  The fact that we detect this source and that it has a brightness close to the expected value confirms that our calibration was successful, and that the non-detection of the HH objects was not due to calibration errors.

\section[]{Discussion}
\subsection{OH Spectra and Maser Emission}
1720-MHz OH masers are an excellent signpost of the interaction of  SNRs with dense molecular material, and have been detected in $\sim10$\%  of the more than 250 SNRs searched (e.g.\ \citeauthor{Yusef-Zadeh1995} \citeyear{Yusef-Zadeh1995}; \citeauthor{Green1997} \citeyear{Green1997}; \citeauthor{Brogan2004} \citeyear{Brogan2004}; \citeauthor{Hewitt2009} \citeyear{Hewitt2009}).  However, only $\sim$ 25\% of the SNRs searched show some signature of interaction  (e.g.\ bright IR, thermal X-rays and broad CO lines), and if we consider only  those  SNRs interacting with molecular material,  the fraction of maser emitting SNRs become much higher. HH objects, on the other hand, often show signatures of interaction with dense molecular material and many optically detected HH objects and jets show shock-excited NIR emission in the lines of molecular hydrogen, observed coincident with, or close by the optically detected shocks.   Although there are cases where  HH objects have broken out of their parent molecular clouds into the ISM and move through material that is mostly atomic, we have selected exclusively objects  for which molecular shocks have been detected. 

We carried out a search for OH maser emission towards 97 HH objects, and  found no conclusive evidence of maser emission in any of them. Thus, despite the similar conditions, SNRs are far more likely to show 1720-MHz OH maser emission than are HH objects. 

The OH maser lines associated with SNRs are distinguished from their  H\footnotesize{\MakeUppercase{\romannumeral 1}\MakeUppercase{\romannumeral 1}}\normalsize{}-region counterparts in having weak emission only at 1720 MHz, relatively low degrees of linear and circular polarisation ($\sim$ 5--10 \%) and appreciable line widths. In single-dish observations the 1720-MHz maser emission shows single line profiles that appear essentially unpolarised with line widths of about 1--2 \kms\ and peak line velocities at or close to the systemic velocity of the remnant. Thus a positive detection of OH emission in low-resolution observations does not necessarily discriminate between weak stimulated radiation from compact maser features and the distributed emission from thermal gas. However,  the difference in the emission and absorption patterns from masers to those generally found from thermal gas in star-forming regions prompted \citet{Goss1968} to suggest that the OH is physically associated with the SNRs. 
In contrast to the situation
in thermal gas, in SNRs,  a narrow emission line is usually seen at 1720~MHz, with broad absorption features (up to 50 \kms) at 1612, 1665 and 1667 MHz.  We did detect 1720 MHz emission for a number of sources in our GBT observations and these do show line widths (0.49--2.87 \kms) comparable to those observed towards maser emitting SNRs (e.g.\ \citeauthor{Hewitt2008} \citeyear{Hewitt2008}). However, 
in our GBT observations, the corresponding 1720-MHz lines appear to be thermal as they are quite weak (30--110 mJy) and the 1665- and 1667-MHz lines are in emission with line widths comparable to those of the 1720-MHz lines.  

For masers in outflows, we would expect the maser emission to originate in the accelerated material, and that in general therefore that the maser velocity would differ from the systematic velocities by amounts up to the outflow velocity. Outflow velocities of up to $\sim$450~\kms\ can be expected from optical jets and their associated HH objects. Our GBT observations, however, showed that the 1720-OH velocities are all within 2~\kms\ of the systemic velocities, suggesting emission not associated with the outflow, and therefore more likely of thermal rather than of maser origin.  Of course if the outflow is in the plane of the sky, the maser velocity would be close to the systemic velocity.  It is however, unlikely that  all of the objects from our observations lie in or close to the plane of the sky. 

The 1720-MHz OH masers in SNRs can only occur under a very restricted set of conditions \citep{Lockett1999}, and different conditions e.g.\  high dust temperatures or a strong
far-ultraviolet (FUV) flux (e.g.\ from a high-mass protostellar source),  could invert the OH main-lines instead and decrease the inversion of the 1720-MHz line. 

\citet{Pashchenko1980} searched for OH main-line masers towards 38 HH objects with the Nancay radio telescope and detected 1667-MHz OH emission in the majority of them, but for most of their sources the 1665-MHz line was below their sensitivity of only a few 10's of mJy.   Apart from one source, their spectra show thermal behaviour with single-peak profiles of 1--3 \kms\ wide, no significant polarisation and for those sources where the 1665-MHz line was detected the main-line intensity ratios are characteristic of LTE.  The one exception, N37 showed apparently strong maser emission, however, \citet{Norris1980}  showed that the maser emission actually came from a nearby H\footnotesize{\MakeUppercase{\romannumeral 1}\MakeUppercase{\romannumeral 1}} \normalsize region and not from the HH object. 
The idea that OH main-line masers can be pumped by bipolar outflows was again suggested by \citet{Argon2003} who detected possible weak OH masers in the W3 OH region which appear to form at the edges of the outflow of a protostellar source, the Turner-Welch object.  From their VLA observations they detect, narrow ($\sim$ 0.3 \kms),  weak ($\sim$200 mJy) main-line emission with the 1665-MHz line stronger than the 1667-MHz line. The emission shows significant linear polarisation with the line  peak velocities consistent with the shift in velocity across the source displayed by thermally excited molecular gas. 

In our GBT observations we detected weak ($0.03-0.81$ Jy), narrow ($0.25-3.14$ \kms) main-line emission in the majority of our sources. However, the  emission lines are unpolarised and at or close to the systemic velocity of the cloud and in most cases the main-line intensity ratios  are between 1 an 1.8 consistent with extended, thermal OH emission associated with dark clouds. 

The non-detection with the VLA of the five sources for which we saw anomalous line emission with the GBT suggest that the emission must be extended. As a reasonable estimate of the largest angular scale visible to the EVLA we calculate the fringe spacing of the shortest baseline in the VLA data for the antennas used, which was $36\arcsec$ at 1.8 GHz. We conclude that the emission from the GBT observations for HH106/107, HH167, HH200 and HH279 must therefore originate from thermal or quasi-thermal excitation processes and is extended on scales $> 36\arcsec$. Spatially extended 1720-MHz OH emission  associated with compact maser spots has been detected toward a large number of maser emitting SNRs. This extended 1720-MHz emission is interpreted  as being of weak maser origin, rather than being thermal \citep[e.g.][]{Yusef-Zadeh1995, Yusef-Zadeh2003, Hewitt2008}, as it shows narrow line-widths and  is  well correlated with OH absorption features at 1612, 1665 and 1667 MHz.

\citet{Hewitt2008} observed thirteen maser emitting SNRs with the GBT and found significantly higher flux densities than had been seen with VLA observations, and they interpret this as evidence of extended maser emission that is resolved out by the VLA  A-array, due to its lack of sensitivity to low brightness temperatures and large spatial scales.
Bright compact maser spots are often only seen when the shock is propagating perpendicular to the line of sight and  extended low gain masers are a result of  weak amplification in the OH behind a face-on shock front, as opposed to the edge-on geometry required by maser spots \citep{Yusef-Zadeh2003}.   Isolated extended maser emission without any compact maser spots has also been detected \citep{Hewitt2008}, and is generally weaker (brightnesses some 20 times lower) than the compact maser emission, with only slightly broader line widths ($\sim 2$ \kms) and low brightness temperatures ($\leq 2500$K), and in some cases at velocities significantly different from the compact masers spots, depending on the extent and location of the emitting region.

Apart from the obvious thermal spectra from our GBT survey towards HH objects, we identified one  source where the 1720 MHz line show very strong and narrow emission compared to the main lines, as well as 10 other sources  with possible non-thermal main-line emission. Our follow-up VLA observations did not show any signs of maser emission. Note, however,   that not all of  our  sources which showed  possible non-thermal emission in the single-dish spectra were re-observed with the VLA, and we cannot of course rule out maser emission in the ones which were not re-observed with the VLA.   However, while we did detect some sources showing anomalous emission, the observed OH emission from our GBT observations is quite likely to be thermal, and while the presence of weak extended masers is not excluded, we have no strong reason to think that any are present. 

\subsection{Physical Conditions Required for Shock-Excited OH Masers}
In order to try and understand why we did not detect 1720-MHz OH masers from our sample of HH objects we need to look  more closely at the conditions needed for the inversion of the 1720-MHz OH line in SNRs, and whether these conditions are in fact likely to occur in HH objects.
 The model developed by \citet{Lockett1999} shows that the masers in SNRs are collisionally excited when the SNR interacts with dense molecular material, but can only occur for a very restricted set of conditions. The physical conditions needed to produce the 1720-MHz OH masers require post-shock temperatures between $T_{K} \sim 50 \-- 125$ K and  $n_{\mathrm{H}_{2}}$ densities between $10^{4}$ and $10^{5}$ cm$^{-3}$. The model also requires OH column densities in the range $N_{\mathrm{OH}} = 10^{16} \-- 10^{17}$ cm$^{-1}$, and such a column must be produced for the shocked material as it cools down to between  50 $\--$ 125 K. \citet{Lockett1999} suggests that the size of this region suitable for maser emission must be at least $\sim 10^{16}$ cm. The model further shows that only  a non-dissociative C-type shock can provide the physical conditions needed to produce the 1720-MHz OH masers, and require a pre-shock density of $n_{\mathrm{H}_{2}} \sim 10^{4}$ cm$^{-3}$, and a shock velocity of  $v_{s} \sim 10 \-- 50$ \kms. In order to get detectable maser amplification these shocks need to be transverse to the line of sight to get the largest velocity coherence.  In addition the OH necessary for the production of 1720-MHz OH masers in SNRs is not directly produced by the shock chemistry \citep{Draine1983}, and an ionization rate of $\zeta \sim 10^{-16}$ s$^{-1}$ is required to produce a sufficient column of OH. \citet{Wardle1999}, concluded that the soft X-ray flux due to bremsstrahlung from the hot gas in maser-emitting SNRs will produce a sufficient column of OH  to allow the formation of the 1720-MHz OH masers.

In dark clouds, the dissociation of many molecular species, in particular that of H$_{2}$O to OH, is primarily due to a FUV flux resulting from energetic electrons produced from ionization of H$_{2}$ by cosmic rays. A typical ionization rate is $\zeta \sim 10^{-17}$ s$^{-1}$. This  is sufficient to explain the enhanced OH/H$_{2}$O abundances associated with the molecular gas in the HH54 outflow \citep{Liseau1996} for example, but is not sufficient to produce the OH  required for the excitation of the 1720-MHz OH masers as in SNRs.   X-ray emission toward a number of HH shocks has been detected with sensitive Chandra and XMM Newton observations \citep[e.g.][]{Pravdo2005}.  The relatively few HH objects with confirmed X-ray detections have  X-ray luminosities  between $\sim 10^{29}$ and $10^{30}$ ergs~s$^{-1}$.  
The X-ray emission detected from these HH objects shows a very soft spectrum and  an extended spatial distribution, similar to X-rays detected from maser emitting SNRs. The X-rays originate from a plasma with a temperature of $\sim 10^{6}$~K, associated with the hottest part of the HH shock, where atomic emission from the J-type shock is seen \citep{Pravdo2001}. 

The X-ray flux decreases as a function of radius from the source of the X-ray emission. 
For maser-emitting SNRs, \citet{Wardle1999} calculated an ionization rate of  $\zeta \sim
10^{-16}$ s$^{-1}$, using a typical X-ray luminosity of $L_{X} \sim 10^{36}$
ergs s$^{-1}$ and a radius of 10~pc. The radius is measured from the centre of the SNR where  the  X-ray emission originates to the edge of the SNR where the 1720-MHz masers are excited.  
For the 5 HH objects with confirmed X-ray detections we  employ the same calculation as \citet{Wardle1999}, but in our case using the X-ray luminosity to determine the radius, away from the apex of the HH bow shock or origin of the X-ray emission, for which  $\zeta$ would be  $\geq10^{-16}$ s$^{-1}$.  
From our calculations we found the radius to be between 0.01~pc ($3\times10^{16}$ cm) and 0.1~pc ($3\times10^{17}$ cm),
depending on the X-ray luminosity of the source. Comparing our calculated values of the radius to the sizes measured for typical bow shaped HH shocks (see discussion below), we find that it is quite likely that the  X-ray emission from the apex of a HH bow shock could have a significant effect on the abundance of  OH in the molecular shocks further away in the wings of the bow.   It is thus important that the effect of these X-rays be included in future hydrodynamic models of HH objects. However, at  the moment  we are dealing with detections where the peak of the X-ray emission detected in HH objects lies around the limit of the capabilities of current instruments. Next generation X-ray instruments could in future detect  X-ray emission from many more HH objects and determine the location and extent of the X-rays.  The details of our
calculations as well as the tabulated X-ray properties of the
individual HH objects are given in Appendix A\@.  

Detailed magneto-hydrodynamic modelling to determine whether the conditions required to produce 1720 MHz OH masers in SNRs  might be present, has only been done for a few HH objects, e.g.\ HH7 \citep{Smith2003}, HH211 \citep{OConnell2005} and  HH240 and 241 \citep{OConnell2004}. These objects all display a bow-shaped structure and the models have all employed the bow model used by \citet{Smith2003}. This bow shock model has a shock with  a paraboloidal geometry where the velocity and excitation temperature both decrease as one moves away from the apex of the bow shock into the wings and as a result the gas behind the shock will decelerate and cool  at a slightly different rate along the shock wings. The bow shock is split into two regions: the apex of the bow described by a dissociative, strong J-type shock (responsible for the atomic emission) and the H$_{2}$ emitting tail described by a C-type shock. The bow shock models that best reproduce the observed optical and NIR emission in the aforementioned HH objects give pre-shock  densities of n$_{\mathrm{H}} = (2.5-8) \times 10^{3}$ cm$^{-3}$,  and bow velocities of, $v_{\mathrm{bow}} = 29-70$ \kms.  These values all agree with conditions required by the SNR maser model, taking into account that the velocity decreases along the wings of the bow.

The HH bow shock model is characterised by a scale length (equal to the
radius of the shock around the focus of  a paraboloid shock).  
In the models of the HH shocks mentioned above, this scale length  is between
$4.2 \times 10^{15}$ and $1.3 \times 10^{16}$ cm. The shock thickness in HH bow shocks is usually much less 
than the scale length, for example,  HH7 where  the total
shock thickness was calculated to be $4.2 \times 10^{14}$ cm, only 0.1 of
the scale length. 

For the population inversion which produces the 1720-MHz line in SNRs the required column of OH must be produced as the shocked material cools down between 125 to 50 K. \citet{Lockett1999} suggest that the size of this region suitable for maser emission must extend over a distance of  $\sim 10^{16}$ cm.

The shocks associated with HH objects are small compared to those in SNRs, and are small even compared to the fraction of the SNR shells from which shock molecular emission is observed and conditions are right for maser formation. The small size of HH shocks not only makes it less likely that the required column length of post-shock material is present,  but also makes it statistically less likely that the shock would interact with regions of molecular material of the appropriate density  than it would be in the case of the larger SNR shells. Note that OH maser emission was detected in the protostellar Turner-Welch object \citep{Argon2003}, but that is associated with a higher-mass star and an outflow of larger spatial extent than is the case for HH objects.

\subsection{Conclusion}

We obtained radio spectra for 97 HH objects.  None of them
showed unambiguous signs of OH maser emission, and none of the five objects that did have anomalous OH emission in the single-dish spectra were detected in our follow-up high-resolution VLA observations. We have no strong reason to believe that the 1720-MHz OH emission in any of our 97 HH-objects is in fact maser emission. 

We show that despite the overall similarities of HH objects to interacting SNRs, the probability of HH objects having just the right conditions to produce 1720-MHz OH masers is low. In the HH objects the size of the shocked region is
much smaller, and the column densities needed for maser emission are
likely not reached.

\newpage
\appendix

\section[]{X-ray Ionization Rates}

We give the X-ray properties of HH objects  with confirmed X-ray emission in Table~\ref{xrays}. For each HH object,  the observed energy range of the detected X-rays in keV, the temperature of the plasma, $T_{\mathrm{X}}$ in MK, the hydrogen column density $N_{\mathrm{H}}$ in cm$^{-2}$, as well as the X-ray luminosity  in units of ergs s$^{-1}$ are listed. To produce the required OH column necessary for 1720-MHz OH masers, the X-rays should be able to dissociate sufficient water molecules behind the C-type shock of the HH object. The X-ray induced ionization rate in the gas should be  $\zeta \sim 10^{-16}$ s$^{-1}$ \citep{Wardle1999}. The X-ray ionization rate is estimated from the total X-ray luminosity of the HH object, provided that the hydrogen column density is $N_{\mathrm{H}} \lesssim 10^{22}$ cm$^{-2}$,  using;

\begin{equation}
\label{x1}
\zeta = N_{\mathrm{e}} \sigma F_{\mathrm{X}}
\end{equation}

where  $N_{\mathrm{e}} \approx 30$ keV$^{-1}$ is the mean number of primary and secondary electrons generated by the absorption of unit energy deposited by X-rays, $\sigma \approx 2.6 \times 10^{-22}$ cm$^{2}$ is the photo-absorption cross section per hydrogen nucleus at 1 keV. 

The X-ray flux, $F_{\mathrm{X}}$,  is the X-ray luminosity at a distance $R$ from the source of the X-ray emission,  given by;

\begin{equation}
\label{x2}
F_{\mathrm{X}} = L_{\mathrm{X}} / 4 \pi R^{2}
\end{equation}

Considering a HH object  with a bow shock geometry, what we want to determine is  if the  X-ray flux produced at the apex of a bow shock would provide a sufficient ionization rate in the wings of the bow shock for production of 1720-MHz OH masers. In order to do this we calculate the distance $R$, from the origin of the X-ray emission, for which the ionization rate $\zeta$ would be   $\geq 10^{-16}$ s$^{-1}$ using equations~\ref{x1} and~\ref{x2} with;

\begin{equation}
\label{x3}
R = \sqrt{L_{\mathrm{X}} / 4 \pi F_{\mathrm{X}}}
\end{equation}

The values of $R$, calculated for each of the
HH objects are also listed in Table~\ref{xrays}. The X-ray
luminosities of the listed HH objects is between $L_{\mathrm{X}} = 1.1
- 450 \times 10^{29}$ ergs s$^{-1}$ and the calculated value of
$R$ between $ \sim 0.01$ and $\sim 0.1$ pc.

\begin{table*}
 \centering
 \begin{minipage}{140mm}
  \caption{HH objects with confirmed X-ray emission.}
  \label{xrays}
  \begin{tabular}{@{}lllllll@{}}
  \hline
   HH    		&       Energy  		&  Luminosity & Temp & Column Density & Radius & References           \\
 object        &       range (keV)	& $L_{\mathrm{X}}$ (ergs s$^{-1}$)       &$T_{\mathrm{X}}$ (MK) 		 &  $N_{\mathrm{H}}$ (cm $^{-2}$) &   $R$ (pc) &     \\

 \hline
HH2		&$0.1 - 2.0$	&$5.2 \times 10^{29}$	&$\sim 1.0$	&$\sim 9.0 \times 10^{20}$	&0.013	&\citet{Pravdo2001}\\
HH80/1	&$0.4 - 2.0$	&$4.5 \times 10^{31}$	&$\sim 1.5$	&$\sim 4.4 \times 10^{21}$	&0.121     	&\citet{Pravdo2004}\\
HH154	&$0.5 - 2.0$	&$3.1 \times 10^{29}$	&$\sim 4.0$	&$\sim 1.4 \times 10^{22}$	&0.010   	&\citet{Bally2003}\\
              	&                    	&                                        	&                  	&                                             	&       	&\citet{Favata2002}\\
HH168	&$0.2 - 1.0$	&$1.1 \times 10^{29}$	&$\sim 5.8$	&$\sim 4.0 \times 10^{21}$	&0.006   	&\citet{Pravdo2005}\\
HH210	&$0.5 - 2.0$	&$1.0 \times 10^{30}$	&$\sim 0.80$	&$\sim 8.0 \times 10^{21}$	&0.018   	&\citet{Grosso2006}\\
HH210	&$0.5 - 2.0$	&$2.0 \times 10^{30}$	&$\sim 0.80$	&$\sim 1.0 \times 10^{22}$	&0.026  	&\citet{Grosso2006}\\
 \hline
\end{tabular}
\end{minipage}
\end{table*}

\bsp

\label{lastpage}

\end{document}